\documentclass[11pt,hyper,letter]{JHEP3}
\usepackage{graphicx,amssymb,amsmath,amsfonts,cite}
\usepackage[english]{babel}
\usepackage{amssymb}
\usepackage{amsfonts}
\usepackage{amsmath}
\usepackage{graphicx}
\usepackage{epsfig}
\usepackage{cite}
\newcommand{\bea}{\begin{eqnarray}}
\newcommand{\eea}{\end{eqnarray}}
\newcommand{\non}{\nonumber \\}
\newcommand{\CR}{\non\cr}
\newcommand{\pa}{\partial}
\newcommand{\ld}{\overleftarrow}
\newcommand{\rd}{\overrightarrow}

\newcommand\be{\begin{equation}}
\newcommand\ee{\end{equation}}

\vspace{18pt}

\title   {\Large  Breaking conformal invariance-\\
     Large N Chern-Simons theory coupled to massive fundamental fermions }
\author
   { Yitzhak Frishman${}^1$\footnotemark[1]  and Jacob Sonnenschein${}^2$\footnotemark[2] \\
${}^1$\textit {
     Department of Particle Physics and Astrophysics\\
The Weizmann Institute of Science, Rehovot 76100, Israel} \\
\\
${}^2$\textit{
 	 The Raymond and Beverly Sackler School of Physics and Astronomy,\\
	Tel Aviv University, Ramat Aviv 69978, Israel}\\

\footnotetext[1]{E-mail address: \email{yitzhak.frishman@weizmann.ac.il}}
\footnotetext[2]{E-mail address: \email{cobi@post.tau.ac.il}}

}

\vspace{18pt}
\abstract{
We analyze the theory of massive fermions in the fundamental representation coupled to a $U(N)$  Chern-Simons gauge theory  at level $K$. It is done  in the large $N$, large $K$ limits where $\lambda=\frac{N}{K}$ is kept fixed.
 Following \cite{Giombi:2011kc}  we  obtain the solution of a Schwinger-Dyson equation for the two point function, the exact expression for  the fermion propagator and the partition function at finite temperature. We prove  that in the large $K$ limit  there exists an infinite set of  classically conserved high spin currents also when a mass is introduced, breaking the conformal invariance. In analogy to the seminal work of 't Hooft on two dimensional QCD,  we write down  a Bethe-Salpeter equation for the wave function of a ``quark anti-quark" bound state. We show that unlike the two dimensional QCD case, the three dimensional Chern-Simons theory does not admit a confining spectrum.}


\preprint{WIS/05/13-MAY-DPPA,  TAUP-2966/13}
\newpage

\begin{document}
\section {Introduction}\label{Intoduction}
In recent years a  major progress has been made in the understanding  of large $N$ three dimensional  Chern-Simons theory coupled to matter in the fundamental representation \cite{Giombi:2011kc}-- \cite{Banerjee:2012gh}. Interesting exact results have been derived without the aid of supersymmetry. Among these achievements is the determination of   the exact planar free energy of the theory at finite temperature on ${\cal R}^2$ as a  function of the 't Hooft coupling $\lambda=\frac{N}{K}$, where $K$ is the level of the Chern-Simons
term. Another property of these theories is the fact that classically in the large $N$ there is an infinite tower of high-spin  conserved currents. It was shown in \cite{Giombi:2011kc} that the divergence of these currents is  equal to a double  and a triple trace  of currents that vanish in the large N limit. In \cite{Aharony:2011jz} it was shown that in the large $N$ limit the theory of $N$ scalars coupled to $U(N)$ CS theory at level $K$ is equivalent to the Legendre transform of the theory of $K$ fermions coupled to a $U(K)$ CS theory at level $N$.


In \cite{Giombi:2011kc} the fact that one can extract exact results is attributed to the discrete nature of the CS  coupling constant, the large $N$ limit, the light-cone gauge and the fact that for the massless case the theory is conformal invariant.
The main question addressed in the  this  work is  to what extent can one decipher  the large N CS theory coupled to massive fundamental fermions. Thus our question is essentially whether two of the three ingredients of the CS coupling, large $N$ and the light-cone gauge are enough to enable us to solve it exactly or is conformal symmetry necessary for that.  Our answer is that there  are interesting physical quantities  that can be determined even without conformal invariance.
 Concretely we have addressed the following  three questions: (i) The fermion propagator and the thermal free energy. (ii) The hight spin currents and their classical conservations. (iii) The spectrum bound state mesons.

Following \cite{Giombi:2011kc} we show that  by solving a Schwinger-Dyson equation, the fermion propagator and the partition function at finite temperature  can be determined exactly. We have generalized the result of \cite{Giombi:2011kc} to the massive case while  using a  somewhat different technique. In \cite{Aharony:2012ns}  it was shown that the result of \cite{Giombi:2011kc} is incomplete and that   there is an additional contribution to the thermal  free energy from   winding modes. The full expression written down in that paper  holds for fermions of any mass,  with an appropriate modification of the parameters.

We prove  that in the large $N$ limit  there exists an infinite set of  classically conserved high spin currents.
The conservation holds classically for high spin currents which are similar to the ones used in the massless case apart from the following replacement
\be
\label{add mass}
(\overleftarrow{D_{\sigma}}\overrightarrow{D^{\sigma}})\rightarrow (\overleftarrow{D_{\sigma}}\overrightarrow{D^{\sigma}})-m^2
\ee
 The divergence of these currents is equal to double trace operators which vanish in the large $N$ limit.  This is the same structure as for the conformal invariant setup.

As for the  spectrum of bound state mesons, we write down,  in analogy to the seminal work of 't Hooft on two dimensional QCD,  a Bethe-Salpeter equation for the wave function of a ``quark anti-quark" bound state. We show that unlike the two dimensional QCD case, the three dimensional Chern-Simons theory does not admit a confining spectrum. In fact, no high mass bound states exist.


The paper is organized as follows: The next section describes  the basic setup of a Chern-Simons theory in Euclidean three dimensions in the large $N$ and large level
 $K$ limits with fixed ratio, coupled to a fermion in the fundamental representation. Section \S 3 is devoted to the determination of the fermion propagator at zero temperature. In section \S 4 we determine the fermion propagator at finite temperature. This is in fact a straightforward generalization  of the result found in \cite{Aharony:2012ns} for the theory which at zero temperature is  conformal invariant. In section \S 5 we discuss the free energy case. Section \S 7 is devoted to the high spin currents.
In section \S 7 we write down a 't Hooft-like equation for the bound states of the theory at zero temperature, and transform it to a form closer to the two dimensional case. We then apply in subsection \S7.1 the high mass approximation, finding a solution that is not consistent with the approximation. The conclusion is that there are no
high mass bound states, where by high mass we mean much larger than the quark mass.
Section \S 8 is devoted to analyzing higher spin currents. We show that the same structure that occurs for the conformal theory is also characterizing the massive theory.
In the last section we summarize our results and present several open questions.

\section{The setup}\label{Ts}
The ${\cal R}^3$  Euclidean action of the $U(N)$ CS theory coupled to a massive  fermion in the fundamental representation is
\be
S=\frac{iK}{4\pi}\int d^3x Tr [A d A + \frac{2}{3} A^3]  + \int d^3x\bar\psi( \gamma^\mu D_\mu + m_{bare}) \psi
\ee
where $A= A^a T^a$, $T^a$ is a fundamental generator normalized so that $Tr[(T^a)^2]=\frac12$ and $D_\mu\psi= \pa_\mu\psi - iA_\mu^aT^a\psi$. Note that we set the coupling constant  to one.
Using light-cone coordinates $x^+,x^-, x^3$ and light-front gauge $A_-=0$ the action in
 momentum space  reads
\bea
S&=& \int\frac{ d^3 p}{(2\pi)^3}\left [ - \frac{iK}{2\pi}Tr[A_3(-p)p_- A_+(p)] +\bar\psi(-p)( i\gamma^\mu p_\mu + m_{bare}) \psi(p)\right ]\CR
&-& i\int\frac{ d^3 p}{(2\pi)^3}\int\frac{ d^3 q}{(2\pi)^3}\left [\bar\psi(-p)[\gamma^+A_+(-q) + \gamma^3 A_3(-q)]\psi(p+q)\right ]\CR
\eea

Here
\bea
x^{\pm}&=&{1\over{\sqrt 2}} [x^1 \pm x^2]\CR
A^{\pm}&=&{1\over{\sqrt 2}} [A^1 \pm A^2]\CR
\eea

It follows from this action that the gauge field propagator takes the form
\be
<A^a_\mu(p) A^b_\nu(-q)> = (2\pi)^3\delta (p-q)\delta^{ab}G_{\mu\nu}(p)
\ee
where the only non-trivial components of $ G_{\mu\nu}(p)  $ are
\be
G_{+3}(p)=-G_{3+}(p)= \frac{4\pi i}{K} \frac{1}{p^+}
\ee
This translates in configuration space to
\be
<A_3(x)A_+(0)>= -<A_+(x)A_3(0)>= \frac{2}{K} \frac{\delta(x_3)}{x^+}
\ee
\section{The fermion propagator}\label{Tfp}

\begin{figure}[h]
\begin{center}
\vspace{3ex}
\includegraphics[width= 100mm]{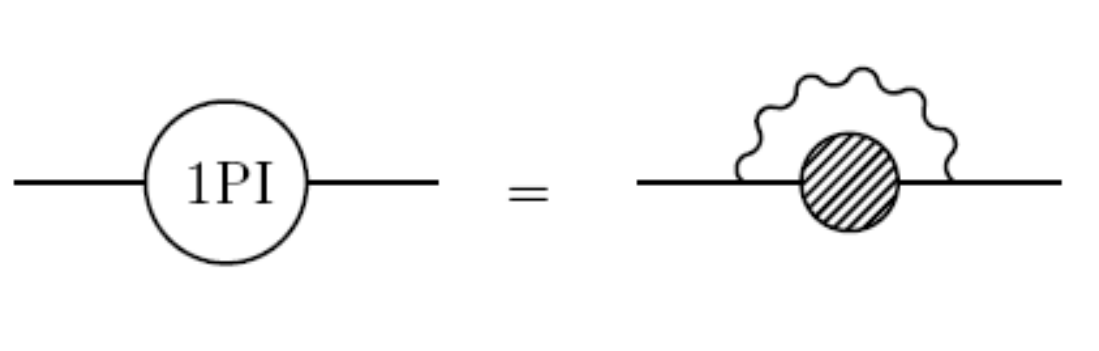}
\end{center}
\caption{Fermion Self Energy}
\label{Self}
\end{figure}

The fermion propagator is given by
\be
<\psi^m(q)\psi_n(-p)>= (2\pi)^3\delta(q-p)\delta^{m}_n S(q) = (2\pi)^3\delta(q-p)\delta^m_n \frac{1}{i\gamma^\mu q_\mu + m_{bare} + \Sigma(q)}
\ee
and
\be\label{sigmaexpand}
\Sigma(q)= i\Sigma_\mu \gamma^\mu +\Sigma_I I  - m_{bare} I
\ee
The equation for $\Sigma$ takes the form
\be\label{FP}
\Sigma(p) = -i 4\pi \lambda\int\frac{ d^3 q}{(2\pi)^3}\frac{\gamma^+ \Sigma_I+ i\emph{I}(q+\Sigma(q))_-}{(q_\mu+\Sigma_\mu(q))(q^\mu+\Sigma^\mu(q)) + \Sigma_\emph{I}(q)^2}\frac{1}{(p-q)^+}
\ee
where $\lambda = N_c/K$.
This is depicted in fig.(\ref{Self}).

Equating the coefficients of the various $\gamma^\mu$ matrices it is clear that
$\Sigma$ is independent of $p_3$ and
\be\label{defgf}
\Sigma_-=\Sigma_3=0 \qquad  \Sigma_I=p_sf_0(\lambda,p_s,m_{bare})\qquad \Sigma_+=p_+g_0(\lambda,p_s,m_{bare})
\ee
with
\be
p_s = \sqrt{p_1^2+p_2^2}= \sqrt{2}|p^-|=\sqrt{2}|p^+|.\ee
Substituting \ref{defgf} into (\ref{FP}) we get the following integral equations for $f_0$ and   $g_0$
\bea\label{g0f0}
g_0  &=& -\frac{4\pi\lambda}{p^-}\int\frac {d^3 q}{(2\pi)^3}\frac{q_s f_0}{q_3^2 + q_s^2(1+ g_0+ f_0^2)}\frac{1}{(p^+-q^+)}\CR
f_0 p_s - m_{bare}  &=& {4\pi\lambda}\int\frac {d^3 q}{(2\pi)^3}\frac{q^+}{q_3^2 + q_s^2(1+ g_0+ f_0^2)}\frac{1}{(p^+-q^+)}\CR
\eea
To solve for the functions $f_0(\lambda,p_s,m_{bare})$ and $f_0(\lambda,p_s,m_{bare})$ we now employ the identity
\be\label{delta}
\frac{\pa}{\pa p^-} \frac{1}{p^+} = 2\pi \delta^2 (p)
\ee
Applying this to Eq (\ref{g0f0}) we get
\bea\label{DE}
\frac{\pa}{\pa p^-} (p_s f_0) &=& \frac {\lambda p^+}{p_s \sqrt{1 + g_0 + f_0^2}} \CR
\frac{\pa}{\pa p^-} (p^- g_0) &=& - \frac { \lambda f_0}{\sqrt{1 + g_0 + f_0^2}}
\eea

Multiplying the first with $f_0 p_s$, the second by $p^+$ and adding, we get zero for the right hand side,
thus obtaining

\be
(1+ \frac{1}{2} p_s \frac {\pa}{\pa p_s})(g_0 + f_0^2) = 0
\ee

Which gives the solution

\be\label{fgrelation}
g_0 + f_0^2 = \frac {m^2}{p_s^2}
\ee

The constant of integration comes out to be $m^2$, where $m$ is the pole in the full propagator.
Using this, the first equation in (\ref{DE})  can be integrated, to give
\be
p_s f_0 = \lambda \sqrt {p_s^2 + m^2} + C
\ee
To determine C, we will evaluate the integral in (\ref{FP}) for $p_s=0$. Actually, it is enough to
evaluate the scalar part. So we have
\be
p_s\rightarrow 0 :(p_s f_0- m_{bare}) \rightarrow  -4\pi \lambda\int\frac{ d^3 q}{(2\pi)^3}\frac{1}{q^2 + m^2}
\ee

The integral is equal to
\be
-\lambda \frac{2}{\pi} \int_0^{\infty} dq + \lambda m
\ee

Setting the linearly divergent integral to zero, by dimensional regularization, we get that $C=m_{bare}$,
and so
\be
p_s f_0 = \lambda \sqrt{p_s^2 + m^2} + m_{bare}
\ee

As for $p_+ g_0$, it follows from the integral equation (\ref{FP}) that it vanishes at $p_s = 0$.
This means that $p_s f_0$ equals $m$ for  $p_s = 0$, entailing
\be\label{mbare}
m_{bare} =  m (1-\lambda)
\ee
Thus the functions  $f_0$ and $g_0$ and hence  the non-trivial components of $\Sigma$  are given by
\bea\label{fg}
p_sf_0(\lambda,p_s,m_{bare})&=& m +\lambda[\sqrt{p_s^2 + m^2} -m] \CR
p_s^2g_0(\lambda,p_s,m_{bare}) &=& -\lambda \left[2m(1-\lambda)[\sqrt{p_s^2 + m^2}-m]+\lambda p_s^2 \right]
\eea

Note that we got this solution without solving for the integrals, just by their form and their values at $p=0$.

It follows from (\ref{fg}) that $\Sigma$ takes the form
\be
\Sigma(p)= ip_+\left [ -\lambda^2 -2\lambda(\frac{m_{bare}}{p_s})(\sqrt{1+\frac{m^2}{p_s^2}}- \frac{m}{p_s}) \right ]\gamma^+ +\lambda p_s\sqrt{1+\frac{m^2}{p_s^2}} I
\ee
Thus the coefficient of the unit matrix in $\Sigma$, which was  for  the massless case $\lambda p_s$, is still linear in $\lambda$ but there is a  re-scaling of $p_s\rightarrow p_s\sqrt{1+\frac{m^2}{p_s^2}}$. The coefficient of  $\gamma^+$ , $i p_+g_0$, which for the massless case was $-ip_+\lambda^2$, is determined  in the massive case  from the relation (\ref{fgrelation}).

It is easy to check that for the massless  limit these results  go back to
\be
f_0= \lambda \qquad g_0 = -\lambda^2
\ee
To summarize the propagator of the massive fermion takes the form
\be
S(q) = -\frac{iq_-\gamma^-+ iq_+(1-g_0)\gamma^+ +iq_3\gamma^3 -f_0 q_s}{q^2+m^2}
\ee
\section{ The fermion propagator at finite temperature}\label{Tfpaft}
The physical system considered in the previous section can be brought into an equilibrium  with a thermal bath of temperature $T=\frac{1}{\beta}$.  To analyze such a case one is required to compactify the Euclidean time direction and thus  put the system   on a  ${\cal R}^2\times S^1$ background manifold rather than  on an ${\cal R}^3$ manifold.
The fermion propagator for such a system  at finite temperature was computed for the massless case in \cite{Giombi:2011kc}. This result was later corrected in  \cite{Aharony:2012ns} where the contribution from  windings around the Euclidean time direction were incorporated and  in addition a mass term was added.
The result derived in \cite{Aharony:2012ns} takes the following form
\be
yf_0(y) -\beta m_{bare}= -\lambda y\sqrt{1+\frac{\mu_T^2}{y^2}} +\frac{1}{\pi i}\left [ Li_2\left ( -e^{-y\sqrt{1+\frac{\mu_T^2}{y^2}}+\pi i\lambda} \right ) -c.c\right ]
\ee
where $y= \beta p_s$ and $Li_2[...]$ is the dilogarithm function.
The  ``thermal mass parameter" $\mu_T$ is determined by the equation
\be
 \mu_T(\lambda)=| \beta m_{bare} +\lambda \mu_T+\frac{1}{\pi i}\left [ Li_2\left ( -e^{-\mu_T-\pi i\lambda} \right ) -c.c\right ]|
\ee
where the positive solution for $\mu_T$ was chosen. At finite temperature there is a non-trivial thermal mass parameter both for the underlying massive as well as the massless theories. In terms of the fermion propagator the difference between the massless and massive theories is the fact that for the former case one has to substitute $m_{bare}=0$ in the equations above.
 Note also  that when $T\rightarrow 0$, and with $\mu_T = \beta m_{T}$, we get that $m_{T} (1-\lambda) \rightarrow m_{bare}$, as expected (use also (\ref{mbare})).

Once the $f(y)$ function is determined we are left over only with $g(y)$ which determines the coefficient of $\gamma^+$ in $\Sigma$. This function is determined in this case also from a relation similar  to  (\ref{fgrelation}), namely
\be
f_0^2(y) + g_0(y) =\frac{ \mu^2_T(\lambda)}{y^2}
\ee
\section{ The free energy}\label{Tfe}
Once the fermion self energy is determined we can also compute the free energy of the system.
In \cite{Aharony:2012ns} it was shown that the free energy of the thermal system is given by
\bea
&&\beta F = - NV_2\int_{-1/2}^{1/2} du \int\frac{d^2 q}{(2\pi)^2}\CR
&&\sum_{n=-\infty}^{\infty} Tr\left [ log\left ( i\tilde q_\mu\gamma^\mu + \Sigma_T + m_{bare}\emph{I}\right ) -\frac12\Sigma_T\frac{1}{i\tilde q_\mu\gamma^\mu + \Sigma_T + m_{bare}}\right ]
\eea
where  we have replaced the trace with an integral over  the uniform spread of eigenvalues and where
$\tilde q_\mu = q_\mu -\frac{2\pi |\lambda| u}{\beta}\delta_{3,\mu}$.

The final result derived in \cite{Aharony:2012ns} takes the form
\bea
&&\beta F= \frac{NV_2}{2\pi \beta^2}\{\frac{\mu_T^3}{3}\left ( 1 - \frac{1}{|\lambda|}\right ) \CR
&&+ \frac{\beta m_{bare} \mu_T^2}{2\lambda} -\frac{({\beta m_{bare}})^3}{6\lambda}+\frac{1}{\pi i \lambda}\int_{\mu_T}^\infty dy y  [ Li_2\left ( -e^{-\mu_T+\pi i\lambda} \right ) -c.c ]\}
\eea

\section{High spin currents}\label{Hsc}
In \cite{Giombi:2011kc} it was shown that the free massless theory admits an infinite set of high spin  currents which are classically conserved. The conserved currents of dimension $s+1$ and spin $s$ have the following structure
\be
J^{(s)}_{\mu_1,...,\mu_s}= \bar\psi \gamma_{\mu_1}\left[ polynomial\ of\ order^{(s-1)}\ of (\ld\pa\ and \  \rd\pa  )   \right]\psi + Symmetrized
\ee
where there is a full symmetrization of all the vector  indices.
The polynomial is determined using a generating function\cite{Giombi:2009wh}
\be
{\cal O}(x,\epsilon)= \sum J^{(s)}_{\mu_1,...,\mu_{s}}\epsilon^{\mu_1}...\epsilon^{\mu_s}=\bar \psi \vec\gamma\cdot\vec \epsilon f(\ld\pa, \rd\pa,\vec\epsilon)\psi
\ee
where
\be
 \label{fuve}
 f(\vec{u}, \vec{v}, \vec{\epsilon}) = \frac{\exp{\left( \vec{u}\cdot \vec \epsilon-\vec{v}\cdot \vec \epsilon \right)} \sinh \sqrt{2 \vec{u}\cdot \vec{v} \vec \epsilon \cdot \vec \epsilon -   4\vec{u}\cdot\vec{\epsilon} \vec{v}\cdot\vec{\epsilon}}}{\sqrt{2 \vec{u}\cdot \vec{v} \vec \epsilon \cdot \vec \epsilon -   4\vec{u}\cdot\vec{\epsilon} \vec{v}\cdot\vec{\epsilon}}}
\ee
By Taylor expanding this function the four first conserved currents are found to be
\bea
\label{exmaslesscur}
 J_{\mu}  & = & \bar{\psi} \gamma_\mu \psi \CR
 J_{\mu_1 \mu_2} & = & \bar{\psi}\gamma_{\mu_1} \left( \overrightarrow{\partial_{\mu_2}} - \overleftarrow{\partial_{\mu_2}} \right) \psi \CR
J_{\mu_1 \mu_2 \mu_3} & = & \frac{1}{6} \bar{\psi}\gamma_{\mu_1} \left( 3 \overleftarrow{\partial_{\mu_2}} \overleftarrow{\partial_{\mu_3}}  - 10 \overleftarrow{\partial_{\mu_2}} \overrightarrow{\partial_{\mu_3}} + 3 \overrightarrow{\partial_{\mu_2}} \overrightarrow{\partial_{\mu_3}} + 2  ( \overleftarrow{\partial_{\sigma}}\overrightarrow{\partial^{\sigma}}) \eta_{{\mu_2}{\mu_3}} \right) \psi \CR
J_{\mu_1 \mu_2 \mu_3 \mu_4} & = & \frac{1}{6} \bar{\psi}\gamma_{\mu_1} \Big( \overleftarrow{\partial_{\mu_2}} \overleftarrow{\partial_{\mu_3}} \overleftarrow{\partial_{\mu_4}}
-7 \overleftarrow{\partial_{\mu_2}} \overleftarrow{\partial_{\mu_3}} \overrightarrow{\partial_{\mu_4}}
+ 7 \overleftarrow{\partial_{\mu_2}} \overrightarrow{\partial_{\mu_3}} \overrightarrow{\partial_{\mu_4}}
- \overrightarrow{\partial_{\mu_2}} \overrightarrow{\partial_{\mu_3}} \overrightarrow{\partial_{\mu_4}} \CR
& \phantom{=} &
+2 (\overleftarrow{\partial_{\sigma}}\overrightarrow{\partial^{\sigma}}) \overleftarrow{\partial_{\mu_2}} \eta_{{\mu_3}{\mu_4}}
- 2
(\overleftarrow{\partial_{\sigma}}\overrightarrow{\partial^{\sigma}}) \overrightarrow{\partial_{\mu_2}} \eta_{{\mu_3}{\mu_4}} \Big ) \psi
\eea
It is easy to check that the first two currents are conserved also for the massive theory.
The higher spin currents are conserved provided we make the following replacement
\be
(\overleftarrow{\pa_{\sigma}}\overrightarrow{\pa^{\sigma}})\rightarrow (\overleftarrow{\pa_{\sigma}}\overrightarrow{\pa^{\sigma}})-m^2
\ee

To show this, we use the generating function, after we make this replacement, applying to it $(\overrightarrow\pa + \overleftarrow\pa)\cdot \frac {\pa}{\pa\vec\epsilon}$.
The contribution of the factor $[\exp{\left( \overrightarrow\pa \cdot \vec \epsilon-\overleftarrow\pa\cdot \vec \epsilon \right)}]$ is
zero also in the massive case, since
\be
(\overrightarrow\pa + \overleftarrow\pa)\cdot ( \overrightarrow\pa -\overleftarrow\pa)=\overrightarrow\Box-\overleftarrow\Box
\ee
which vanishes when taken between $\bar\psi$ and $\psi$.
As for the other factors in the generating function, it is enough to show that the contribution of the argument vanishes, namely
\be
2(\overrightarrow\pa\cdot \overleftarrow\pa -m^2) \vec \epsilon \cdot \vec \epsilon -   4\overrightarrow\pa\cdot\vec\epsilon \overleftarrow\pa\cdot\vec\epsilon =0.
\ee
This is indeed the case, as
\bea
&&4(\overrightarrow\pa\cdot \overleftarrow\pa -m^2) \vec \epsilon \cdot(\overrightarrow\pa + \overleftarrow\pa)  -   4\overrightarrow\pa\cdot(\overrightarrow\pa + \overleftarrow\pa)\overleftarrow\pa\cdot\vec\epsilon  -   4\overrightarrow\pa\cdot\vec\epsilon \overleftarrow\pa\cdot(\overrightarrow\pa + \overleftarrow\pa)
\CR
&&= -4(\overleftarrow\pa\cdot \vec \epsilon)(\overrightarrow\Box+m^2) -4(\overrightarrow\pa\cdot \vec \epsilon)(\overleftarrow\Box+m^2)
\eea
which also vanishes when taken between $\bar\psi$ and $\psi$.

It remains to check the part where the divergence is contracted with the $\gamma$ matrix, which follows from
\be
\bar\psi(\overrightarrow\pa_\mu + \overleftarrow\pa_\mu)\gamma_\mu\psi=0
\ee
also in the massive case.

For the interacting massless theory  it was proven in \cite{Giombi:2011kc} that there is a similar set of higher spin currents,  gotten by replacing ordinary derivatives with covariant derivatives,
\be
\label{pa to D}
\ld\pa \rightarrow \ld D \qquad \rd\pa \rightarrow \rd D
\ee
The similarity is in the sense that although not conserved, their divergence does not have a single trace part.

The classical divergence of the corresponding currents was derived using the equation of motion.
The one associated with the gauge fields does not change in form when a mass is introduced, and it remains as
\be
\label{cseom}
(F_{\mu \nu})^i_j=\frac{ \pi}{K} \epsilon_{\mu \nu \rho}
{\bar \psi}^i \gamma^\rho \psi_j
\ee
However, the equation associated with the fermion does of course change, and it now reads
\be
\label{dirac}
D^\mu \gamma_\mu \psi = i m \psi
\ee

Now, when checking the equation for the divergence of the currents, we have to change orders of covariant derivatives.
For this we use
\bea
&&\overrightarrow{D}_\mu \overrightarrow{D}_\nu - \overrightarrow{D}_\nu \overrightarrow{D}_\mu = -i F_{\mu\nu} \CR
&&\overleftarrow{D}_\mu\overleftarrow{D}_\nu - \overleftarrow{D}_\nu \overleftarrow{D}_\mu = i F_{\mu\nu}  \CR
&&\overrightarrow{D}_\mu\overleftarrow{D}_\nu - \overleftarrow{D}_\nu \overrightarrow{D}_\mu = i F_{\mu\nu}
\eea
Combining with \eqref{cseom}, we see that for the single trace terms, we can treat the covariant derivatives as commuting, since the
extra terms are higher trace, with each  one multiplied by an extra $\frac{1}{K}$ factor.

We now claim that also in the massive case, the divergence of the currents will have no single trace term, provided we make the
replacement
\be
(\overleftarrow{D_{\sigma}}\overrightarrow{D^{\sigma}})\rightarrow (\overleftarrow{D_{\sigma}}\overrightarrow{D^{\sigma}})-m^2
\ee
The currents are now

\be
\tilde{J}^{(s)}_{\mu_1,...,\mu_s}= \bar\psi \gamma_{\mu_1}\left[ polynomial\ of\ order^{(s-1)}\ of (\ld D\ and \  \rd D  )   \right]\psi + Symmetrized
\ee
In the following we will omit the "Symmetrized".

Take the divergence
\bea
&&\pa_{\mu_l} \tilde{J}^{(s)}_{\mu_1,...,\mu_s} = \CR
&&\bar\psi(\ld\pa_{\mu_l}+\rd\pa_{\mu_l})\gamma_{\mu_1}\left[ polynomial\ of\ order^{(s-1)}\ of (\ld D\ and \  \rd D  )   \right]\psi =\CR
&&\bar\psi(\ld D_{\mu_l}+\rd D_{\mu_l})\gamma_{\mu_1}\left[ polynomial\ of\ order^{(s-1)}\ of (\ld D\ and \  \rd D  )   \right]\psi
\eea

Now, when $\mu_l$ is contracted with $\gamma$, we have
\bea
&&\bar\psi(\ld D_{\mu_l}+\rd D_{\mu_l})\gamma_{\mu_l}\left[ polynomial\ of\ order^{(s-1)}\ of (\ld D\ and \  \rd D  )   \right]\psi =\CR
&&\bar\psi(i m +\rd D_{\mu_l}\gamma_{\mu_l})\left[ polynomial\ of\ order^{(s-1)}\ of (\ld D\ and \  \rd D  )   \right]\psi=\CR
&&\bar\psi\left[ polynomial\ of\ order^{(s-1)}\ of (\ld D\ and \  \rd D  )   \right](i m +\rd D_{\mu_l}\gamma_{\mu_l})\psi +\CR
&&Terms \ with \ double \ trace \ multiplied \ by \ \frac{1}{K} =\CR
&&Only \ terms \ with \ double \ trace \ multiplied \ by \ \frac{1}{K}
\eea

For the rest of the proof, we use the generating function, with the replacement \eqref{pa to D} to covariant derivatives, and
also the $m^2$ shift  \eqref{add mass}.

As for the term
\be
\exp{\left( \overrightarrow{D}\cdot \vec \epsilon-\overleftarrow{D}\cdot \vec \epsilon \right)}\ee
it is sufficient to check
\be(\overrightarrow{D} + \overleftarrow{D})\cdot ( \overrightarrow{D} -\overleftarrow{D})\ee
which, for the single trace terms, is equal to
\be
\overrightarrow{D}\cdot\overrightarrow{D}-\overleftarrow{D}\cdot\overleftarrow{D}
\ee
This is actually zero, by  the use of
\bea
\label{DD}
&&\overrightarrow{D}\cdot\overrightarrow{D}\psi=-(\frac{1}{2} \epsilon^{ijk}F_{ij}\gamma_k + m^2)\psi \CR
&&\bar\psi \overleftarrow{D}\cdot\overleftarrow{D}=-\bar\psi(\frac{1}{2} \epsilon^{ijk}F_{ij}\gamma_k + m^2)
\eea
which follows from \eqref{dirac}.

It remains to examine the contribution of the factor
\be
2(\overrightarrow{D}\cdot \overleftarrow{D} -m^2) \vec \epsilon \cdot \vec \epsilon -   4\overrightarrow{D}\cdot\vec\epsilon \overleftarrow{D}\cdot\vec\epsilon
\ee

Its contribution to the divergence, for the single trace terms, is
\bea
&&4(\overrightarrow{D}\cdot \overleftarrow{D} -m^2) \vec \epsilon \cdot(\overrightarrow{D} + \overleftarrow{D})  -   4\overrightarrow{D}\cdot(\overrightarrow{D} + \overleftarrow{D})\overleftarrow{D}\cdot\vec\epsilon  -   4\overrightarrow{D}\cdot\vec\epsilon \overleftarrow{D}\cdot(\overrightarrow{D} + \overleftarrow{D})
\CR
&&= -4(\overleftarrow{D}\cdot \vec \epsilon)(\overrightarrow{D}\cdot\overrightarrow{D}+m^2) -4(\overrightarrow{D}\cdot \vec \epsilon)(\overleftarrow{D}\cdot\overleftarrow{D}+m^2)
\eea

It has no single trace terms, by \eqref{DD}. This completes our proof.
\section{'t Hooft like equation for the spectrum of bound states}\label{tHlwftsobs}
In the conformal setup when the fermions are massless a natural question to address is the spectrum of dimensions of the   primaries operators  and  their descendants. The primaries are the operator $\bar \psi \psi$ and the tower of symmetric traceless currents $J^{(s)}_{\mu_1,...\mu_s}$ which are   constructed  from a fermion  anti-fermion bilinear sandwiching  derivatives and a gamma matrix.  The analysis of the spectrum of dimensions was carried out  in \cite{Giombi:2011kc}. The analogous question for the massive theory is the mass spectrum of bound states. The latter can be built in the same way as in the conformal theory.  Here we will discuss a special class of the mesonic bound states.   Note also that since the theory is invariant under  local  $U(N)$ symmetry, and not only $SU(N)$, baryons bound states are not gauge invariant.
\footnote {Note, however, that at large $N$ the $U(1)$ part is down by $1\over N$.}
We address the question of the spectrum of masses only at zero temperature.

\begin{figure}[h]
\begin{center}
\vspace{3ex}
\includegraphics[width= 100mm]{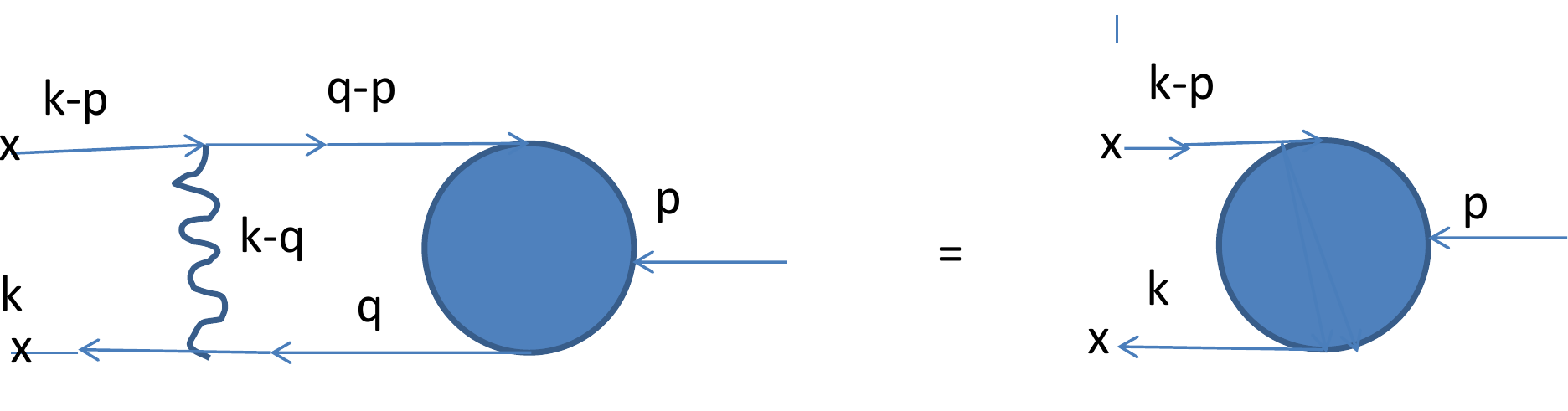}
\end{center}
\caption{Schwinger-Dyson equation for the fermion anti-fermion bound state}
\label{BethSal}
\end{figure}

The spectrum of fermion anti-fermion  bound states of two dimensional QCD in the planar limit was solved in the seminal work of 't Hooft  \cite{'tHooft:1974hx}. Since like in that work, here we are also using (i) light-front coordinates, (ii) light-cone gauge and (iii) the planar limit, it calls for the use of a similar approach to the one used in \cite{'tHooft:1974hx}  for our system. The key player is the bound state ``wave-function" or the ``blob" which is the Fourier transform of the matrix element of the operator $\psi(x) \bar \psi(0)$ between the vacuum and the meson states,
\be\label{phi}
\phi(p,k)= \int \frac{d^3x}{(2\pi)^3}e^{ikx} <meson(p)|T \psi(x) \bar \psi(0)|0>
\ee
To determine the ``wave-function" one has to solve  a  Bethe-Salpeter   which is depicted in figure  (\ref{BethSal}).

 Note that the correlator in the definition of $\phi(p,k)$ includes the operators $\psi$ and $\bar \psi$  at different points $\psi(x) \bar \psi(0)$. Expanding $\psi(x)$ around $x=0$ we get bilinear operators of $\psi$ and $\bar \psi$ with any arbitrary number of derivatives
$(\pa_{\mu_1}...\pa_{\mu_n}\psi)  \bar \psi |_{x=0}$. Thus the blob describes  a bound state of a quark and an anti-quark with all orbital momenta. As for the internal spin, $\phi(p,k)$ is a 2x2 matrix, so it includes the spin zero and one components, and those are all the Dirac bilinear combinations in 3 dimensions.

To determine the masses of the bound states, we will have to go back to Minkowski space.
But let us first continue in Euclidean space.

The integral equation reads
\be\label{Int1}
 \phi(k,p)= -\frac{\lambda}{2\pi^2} S(k) \int\frac{d^3q}{(k^+ -q^+)} \left[ \gamma^+ \phi(q,p) \gamma^3 - \gamma^3 \phi(q,p) \gamma^+ \right ] S(k-p)
\ee

Using  (\ref{delta})  we can transform the integral equation into the following  differential equation
\be
\frac{\pa}{\pa k^-} \left[ S^{-1}(k) \phi(k,p) S^{-1}(k-p) \right] = \frac{\lambda}{\pi} \left[ \gamma^3\tilde\phi(\tilde k,p)\gamma^+ - \gamma^+ \tilde\phi(\tilde k,p) \gamma^3 \right ]
\ee
where $\tilde k$ is the vector $(k_1,k_2)$, and
\be
\tilde \phi(\tilde k,p) = \int d k^3\phi(p,k)
\ee

Next we expand the blob in terms of the coefficients of $\gamma^\mu$ and $\emph{I} $  similar  to (\ref{sigmaexpand})
\be
\phi= \phi_-\gamma^- + \phi_+\gamma^+ + \phi_3\gamma^3 + \phi_s \emph{I}
\ee
which gives
\be
\gamma^3 \phi \gamma^+ - \gamma^+ \phi \gamma^3 = 2[\phi_s \gamma^+ - \phi_-\emph{I}]
\ee
and similarly for $\tilde\phi$.

Thus the right hand side of (\ref{Int1}) involves $\phi_s$ and $\phi_-$ only.
This results in two coupled integral equations for $\phi_s$ and $\phi_-$, with $\phi_+$ and $\phi_3$ determined from $\phi_s$ and $\phi_-$.

The  integral equation now implies
\bea\label{ineqfsfo}
\frac{\pi^2}{\lambda}\phi_s (k,p) &=& [a_-(b_s-b_3)+ b_-(a_3+a_s)]\int\frac{d^3q}{(k^+ -q^+)}\phi_s(q,p)\CR
&-&  [a_-b_++a_+b_- +a_3b_3+ a_s b_s]\int\frac{d^3q}{(k^+ -q^+)}\phi_-(q,p)\CR
\frac{\pi^2}{\lambda}\phi_- (k,p) &=&[2 a_-b_-]\int\frac{d^3q}{(k^+ -q^+)}\phi_s(q,p)\CR
&-&  [ a_-(b_3 + b_s)+b_-(a_s-a_3)]\int\frac{d^3q}{(k^+ -q^+)}\phi_-(q,p)\CR\eea
where
\be
a_s =-\frac{f_0 k_s}{k^2+m^2} \qquad a_-=\frac{ik_-}{k^2+m^2}\qquad a_+ =\frac{ik_+(1-g_0)}{k^2+m^2}\qquad a_3= \frac{ik_3}{k^2+m^2}
\ee
and $(b_s,b_-, b_+, b_3)$ are given similarly with the same expressions but with $k-p$ replacing $k$.

Let us choose now the frame
\be
p = (0, 0, p_3)
\ee

Then
\bea
 a_- &=& \frac{i k_-}{[k^2 + m^2]}  \qquad \ \ \ \ \ \ \
 b_- = \frac{i k_-}{[k_s^2 + (k_3 - p_3)^2 + m^2]}  \CR
 a_+ &=& \frac{i k_+[1 - g_0(k_s)]}{[k^2 + m^2]}  \qquad
 b_+ = \frac{i k_+[1 - g_0(k_s)]}{[k_s^2 + (k_3 - p_3)^2 + m^2]}\CR
 a_s &=& -\frac{k_s f_0(k_s)}{k^2 + m^2}  \qquad \ \ \ \ \ \
 b_s = -\frac{k_s f_0(k_s)}{k_s^2 + (k_3 - p_3)^2 + m^2}  \CR
 a_3 &=& \frac{i k_3}{k^2 + m^2}  \qquad \ \ \ \ \ \ \ \
 b_3 = \frac{i (k_3 - p_3)}{k_s^2 + (k_3 - p_3)^2 + m^2}  \CR
\eea

The integral equations become
\bea
\frac{\pi^2}{\lambda}\phi_s (k,p) &=& -\frac{k^+[p_3 + 2i k_s f_0(k_s)]}{[k^2+m^2][k_s^2+(k_3 - p_3)^2+m^2]}\int\frac{d^3q}{(k^+ -q^+)}\phi_s(q,p)\CR
&+&  \frac{k_s^2 + k_3(k_3-p_3)-m^2}{[k^2+m^2][k_s^2+(k_3 - p_3)^2+m^2]}\int\frac{d^3q}{(k^+ -q^+)}\phi_-(q,p)\CR
\frac{\pi^2}{\lambda}\phi_- (k,p) &=& -\frac{(k^+)^2}{[k^2+m^2][k_s^2+(k_3 - p_3)^2+m^2]}\int\frac{d^3q}{(k^+ -q^+)}\phi_s(q,p)\CR
&+&\frac{k^+[2i k_s f_0(k_s)-p_3]}{[k^2+m^2][k_s^2+(k_3 - p_3)^2+m^2]}\int\frac{d^3q}{(k^+ -q^+)}\phi_-(q,p)\eea

We can now perform an integration over $k_3$ on both sides, noting that  $\int d^3 q$ is independent of $k_3$. This will result in integral
equations for $\tilde \phi$ on both sides.

To find the bound states, we have to go to Minkowski space, by analytic continuation to $p_3=i M_b$.
The solutions of the integral equation should provide us with the masses of the bound states $M_b$.

To perform the integrals over $k_3$, we will make use of the following   integrals
\bea
&&\int \frac{dk_3}{[k_s^2 + k_3^2 + m^2][k_s^2 + (k_3 - p_3)^2 + m^2]} = \frac{2\pi}{\sqrt{k_s^2 + m^2}} \frac {1}{[p_3^2 + 4(k_s^2 + m^2)]} \CR
&&\int dk_3\frac{k_3(k_3-p_3)}{[k_s^2 + k_3^2 + m^2][k_s^2 + (k_3 - p_3)^2 + m^2]} = \frac{2\pi\sqrt{k_s^2 + m^2}}{[p_3^2 + 4(k_s^2 + m^2)]}
\eea

The integral equations, after the $k_3$ integration, become
\bea
& & \frac{\pi}{2\lambda k^+} \sqrt{k_s^2 + m^2} [p_3^2 + 4(k_s^2 + m^2)]\tilde \phi_s (\tilde k,p_3) = -[2i k_s f_0(k_s)+p_3]\int\frac{d^2\tilde q}{(k^+ -q^+)}
\tilde\phi_s(\tilde q,p_3)\CR
&+&  4k^-\int\frac{d^2\tilde q}{(k^+ -q^+)}\tilde \phi_-(\tilde q,p_3)\CR
& & \frac{\pi}{2\lambda k^+} \sqrt{k_s^2 + m^2} [p_3^2 + 4(k_s^2 + m^2)]\tilde \phi_- (\tilde k,p_3) = -k^+\int\frac{d^2 \tilde q}{(k^+ -q^+)}\tilde \phi_s(\tilde q,p_3)\CR
&+&[2i k_s f_0(k_s)-p_3]\int\frac{d^2\tilde q}{(k^+ -q^+)}\tilde \phi_-(\tilde q,p_3)\CR\eea

 The wave functions above may have gauge dependent parts in them.
However, we will be looking at the masses of the bound states, which depend on the gauge invariant parts only.

\subsection{Large bound state mass approximations}
The analysis of the spectrum of bound sates  of large $N$ QCD in two space-time dimensions simplifies in the limit of high excitation, or large bound state mass. It was found out \cite{'tHooft:1974hx} that for high excitations the spectrum becomes  a Regge-like spectrum, namely, $M^2_b\sim n$ where $n$ is the excitation number. Thus, we would like to  first to investigate whether there is a region of large bound sate masses and if yes what is the structure of the spectrum in that region.

Take, in the last equation, the limit of $|ip_3| \Rightarrow \infty$.
Then,
\be\label{IntEq}
\frac{\pi p_3}{2\lambda k_-} \sqrt{k_s^2 + m^2}\tilde \phi_s (\tilde k,p_3) = -\int\frac{d^2\tilde q}{(k^+ -q^+)}
\tilde\phi_s(\tilde q,p_3)
\ee
and an identical one for $\tilde \phi_-(\tilde q,p_3)$.
Define
\be
\psi(\tilde k,p_3)=\sqrt{k_s^2 + m^2}\tilde \phi_s (\tilde k,p_3)
\ee
Then
\be
\frac{\pa}{\pa k^-}\ln\psi(\tilde k,p_3)= -\frac{4\lambda}{p_3} \frac{k^+}{\sqrt{k_s^2 + m^2}}
\ee
from which
\be
\ln\psi(\tilde k,p_3)=-\frac{4\lambda}{p_3} \sqrt{k_s^2 + m^2} + F
\ee

The additional function F, by rotational invariance in the (1,2) plane, depends on $k_s$ only, and as it does not depend on $k^-$,
it therefore is independent of $\tilde k$ altogether. It can thus depend on $p_3$ only.
Finally, we get
\be
\psi(\tilde k, M_b)=\exp{[F(M_b)]}\exp{[i\frac{\lambda}{M_b}4\sqrt{k_s^2 + m^2}]}
\ee
where the proportionality factor may depend on $M_b$. Note that factor $4\sqrt{k_s^2 + m^2}$ is the energy of the relative motion, in the rest frame
of the bound state. Also, since the equations are homogeneous, we might as well put the proportionality factor to 1.

Substituting the solution into the integral on the right hand side of (\ref{IntEq}), we actually get
$$\exp{[F(M_b)]}\{\exp{[i\frac{\lambda}{M_b}4\sqrt{k_s^2 + m^2}]}-\exp{[i\frac{\lambda}{M_b}4m]}\}$$
which means we have no solution to the integral equation (\ref{IntEq}), except the trivial one $\tilde \phi_s (\tilde k,M_b)=0$.

Turning now to $\tilde \phi_- (\tilde k,M_b)$, we have here the form $k_-$ times a function of $k_s^2$.
Looking for a solution for an equation like (\ref{IntEq}), we do find that
$$\sqrt{k_s^2 + m^2}\phi_-(\tilde k, M_b)=k_- G(M_b)\exp{[i\frac{\lambda}{M_b}4\sqrt{k_s^2 + m^2}]}$$
is a solution, provided that we take
$$lim_{k_s^2\rightarrow\infty}\exp{[i\frac{\lambda}{M_b}4\sqrt{k_s^2 + m^2}]}=0$$
in view of the oscillatory factor.

To fix the eigenvalues of $M_b$, we need some boundary conditions.
Assuming, as an example, the requirement that the solution is real at $k_s=0$, we get for the n-th state,
$$M_b^n=\frac{4\lambda m}{\pi n}$$
which means that the highest mass is actually the one for n=1, and it is not extending to $\infty$, contradicting our initial assumption.

It is thus clear that unlike  spectrum of two dimensional QCD, in the large $N$ large $K$ limit of the CS theory coupled to fermions in the fundamental representation the spectrum does not admit a confining behavior. This should not surprise us since we are not considering for the gauge fields the YM theory but rather the CS one.
\section{ Summary and open questions}\label{Saoq}
Two main questions have been addressed in this note:

(i) In the CS theory coupled  to fermions in the fundamental representation, in the large $N$ large level limit,
is conformal invariance really necessary for:

 1. Having   high-spin currents which are classically conserved.

 2. For  the exact determination of the fermion propagator and the thermal free energy.

(ii) The structure of the spectrum of bound states of the massive  theory.

As for the fermion propagator and the thermal free energy we have generalized the results of \cite{Giombi:2011kc} to the case of massive fermions. While working on this project these properties were re-derived including a correction due to the presence of modes associated with the holonomy around the Euclidean circle in \cite{Aharony:2012ns}. These results were determined for both the massless and massive cases.
We have shown in this paper  that the same structure of  classically conserved high spin currents of the massless theory occurs also for the massive one. This was derived by  a simple modification of the currents.

The question of the spectrum of bound states of the three dimensional CS theory is less tractable than the analogous problem in two dimensional QCD\cite{'tHooft:1974hx}. For once there are two coupled integral  equations for two independent components of the `` wave function" $\phi_s(k,p)$ and $\phi_-(k,p)$. The fact that the integrals are three dimensional rather than just two light-cone dimensions add further complication. We were not able to solve the equations in full generality. However, we were able to show that,  unlike the two dimensional case where there is for large excitation number a `` Regge spectrum'' where $[M_b^{(n)}]^2\sim n$, there are no highly excited bound states in our case. This is in accordance with the fact that the system described in this paper is not a confining one as can be seen from (\ref{cseom}).

The research of the  massive CS theory is still in an infant state and there are plenty of open quesitons to further investigate.
\begin{itemize}
\item
For the conformal theories both with fermions and with scalars the high spin currents were analyzed also beyond the classical limit.
In particular three point functions involving high spin currents have been computed\cite{Giombi:2011kc}\cite{Aharony:2012nh}. A natural question to ask is can one determine the analogous correlators for the massive theory.
\item
By no means  the investigation of the coupled `t Hooft like equations has been exhausted. One can try solving them in other special cases similar to the $|ip_3|\rightarrow\infty $ that we have used, and one may try to use also  numerical methods.
\item
Following the work of `t Hooft, a similar though non-homogeneous, integral equation was written down for the scattering amplitude of a quark anti-quark \cite{Callan:1975ps}. In a similar manner to the coupled equations that we derived for the ``blob", one can write equations also for the scattering amplitude and analyze them.
\item
We have addressed the issue of mesonic bound-states at zero temperature. Provided that there are such bound-states at zero temperature an interesting question is to examine their fate once a temperature is turned on.
\item
 In \cite{Aharony:2012nh} a bosonization duality was proposed that relates a certain CS theory coupled to massive fermions in the fundamental representation to a CS theory couple to complex scalar in the fundamental representation. Thus one can investigate the issue of the spectrum of the bounds states of the massive theory also in the bosonic theory.
\item
An interesting limit to address is that of $m_{bare}\rightarrow\infty$ where the theory is supposed to flow to a ``topological" pure CS theory.
\end{itemize}

{\bf Acknowledgements}

We would like to thank Ran Jacobi and Guy Gur-Ari for useful discussions and Ofer Aharony and Adam Schwimmer for comments on the manuscript.  The work of J.S is partially supported by the Israel Science Foundation (grant 1665/10) and by the ``Einstein Center of Theoretical Physics " at  the Weizmann Institute.

\end{document}